\documentclass[12pt]{article}
\usepackage{amsmath,amssymb}
\usepackage{epsfig,epsf}
\usepackage{graphicx, subfigure}
\usepackage{graphics}
\newcommand{\be}{\begin{eqnarray}}
\newcommand{\ee}{\end{eqnarray}}
\newcommand{\nn}{\nonumber \\}
\newcommand{\lb}{\label}
\newcommand{\p}[1]{(\ref{#1})}
\newcommand{\vecg}[1]{\mbox{\boldmath $#1$}}
\newcommand{\vecb}[1]{{\bf #1}}
\begin{document}

\begin{titlepage}
 \begin{center}

{\Large Multidimensional Dirac strings and 
the Witten index of SYMCS theories with groups
of higher rank}

\vspace{1cm}

A.~V. Smilga

\vspace{1cm}

{SUBATECH, Universit\'e de Nantes, \\
4 rue Alfred Kastler, BP 20722, Nantes 44307, France}\\
E-mail: smilga@subatech.in2p3.fr
  \end{center}

\vspace{2cm}

\begin{abstract}
We discuss generalized Dirac strings associated with a given Lie group. They live on ${\mathcal C}^r$
($r$ being the rank of the group). Such strings show up in the effective Born-Oppenheimer Hamiltonian for
$3d$ supersymmetric Yang-Mills-Chern-Simons theories,
 brought up by the gluon loops. 
We calculate accurately the number of the vacuum states in the effective 
Hamiltonian associated with  these
strings. We also show that these states are 
{\it irrelevant} for the final SYMCS vacuum counting. The Witten
index of SYMCS theories depends thus 
only on the strings generated by fermion loops and 
carrying fractional generalized fluxes.

\end{abstract}

\end{titlepage}

\section{Introduction}

\def\theequation{\arabic{section}.\arabic{equation}} 
\setcounter{equation}0

The Lagrangian of pure $3d$ ${\cal N} = 1$ 
supersymmetric Yang-Mills-Chern-Simons theory  
reads
 \be
 \label{LN1}
  {\cal L} \ =\ \frac 1{g^2}  \left \langle - \frac 12 F_{\mu\nu}^2 +
  i\bar \psi /\!\!\!\!D \psi \right \rangle +
  \kappa   \left\langle \epsilon^{\mu\nu\rho}
  \left( A_\mu \partial_\nu A_\rho - \frac {2i}3 A_\mu A_\nu A_\rho \right ) - 
\bar \psi \psi \right \rangle \, .
   \ee
The conventions are:  $\epsilon^{012} = 1, \ D_\mu {\cal O}  = \partial_\mu {\cal O}  - 
i[A_\mu, {\cal O}] $ (such that $A_\mu$ is Hermitian),  
   $\psi_\alpha$ is a 2-component Majorana $3d$ spinor belonging to the adjoint 
representation of the gauge group, and $\langle \ldots \rangle$ stands for the color trace. 
$\gamma$ - matrices can be chosen as
$   \gamma^0  = \sigma^2,\ \gamma^1 = i\sigma^1,\  
\gamma^2 = i\sigma^3 $ \, .

 This is  a $3d$ theory and the gauge coupling constant $g^2$ carries the dimension of mass.
The requirement for the functional integral to be invariant under certain large 
(non-contractible)
gauge transformations (see e.g. Ref.\cite{Dunne} for a nice review)
 leads to the quantization condition
  \be
    \label{quantkap}
    \kappa =  \frac k {4\pi}  \ .
    \ee
For the unitary $SU(N)$ gauge groups, the {\it level} $k$ must be integer if $N$ is even and half-integer  if $N$ is odd.
For an arbitrary group, this depends on whether the adjoint Casimir eigenvalue $c_V$ is even or odd.  

The index of this theory 
\be
\label{Witind}
I \ =\ {\rm Tr} \{ (-1)^F e^{-\beta H} \} 
 \ee
 was evaluated in \cite{Wit99} with the result
 \be
\label{IkN}
I(k,N) \ =\  
[{\rm sgn}(k)]^{N-1} \left( \begin{array}{c} |k|+N/2 -1 \\ N-1 \end{array} \right)\ .
 \ee
for $SU(N)$ gauge group. This is valid for $|k| \geq N/2$. For $|k| < N/2$, the index vanishes
and supersymmetry is broken. In the simplest $SU(2)$ case, the index is just
 \be
\label{Ik2}
I(k,2) \ =\ k \ .
 \ee
For $SU(3)$, it is 
\be
\label{Ik3}
I(k,3) \ =\ \frac {k^2 - 1/4}2 \, .
 \ee
The result (\ref{IkN}) was derived in \cite{Wit99} by  considering the theory in a  large
spatial volume, $g^2L \gg 1$ where the vacuum dynamics is determined by the Chern-Simons  
term with the  coupling renormalized due {\it only} to fermion loops 
\footnote{Note that the renormalized level is always integer.}
  \be
\label{kren}
 k \ \to \ k- \frac N2 \, .
 \ee
The number (\ref{IkN}) is nothing but the full number of states in the topological pure CS theory with
the renormalized level \p{kren}.

For an arbitrary group, the general recipe is
  \be
\lb{recipe}
 I^{\rm full}_{\rm SYMCS}\, (k) \ =\  I^{\rm tree}_{\rm SYMCS} \left( k - \frac {c_V}2 \right)\, , 
  \ee 
where $c_V$ is the adjoint Casimir eigenvalue.

 In Refs.\cite{ja1,ja2} (see also the review \cite{ja3}), 
the result \p{IkN} was reproduced with another method \cite{Wit82}
by  considering the theory in a {\it small} spatial box, $g^2L \ll 1$, and 
  studying
 the dynamics of the corresponding Born-Oppenheimer Hamiltonian. 
(We also evaluated the index for $Sp(2r)$ and $G_2$, 
see Eqs. \p{ISp2r}, \p{indG2}. )
  
If imposing the periodic boundary conditions for all fields, 
the slow variables in the effective BO Hamiltonian are just the zero 
Fourier modes
of the spatial components of the {\it Abelian} vector potential 
(belonging to the Cartan subalgebra) $C_{j = 1,2}^{a = 1,\ldots, r}$ 
 and its superpartners $\psi^a = \psi_{1-i2}^a$,
  The motion in the field space $\{ {C}_j^a \} \equiv \{\vecb{C}_j\}$ is actually 
finite because the shifts 
  \be
\label{shift}
\vecb{C}_j \to \vecb{C}_j + 4\pi {n}_j \vecb {a} /L \, ,
 \ee
where $\vecb{a}$ are the coroots of the group and  $n_j$  are integer numbers, 
 amount to   contractible  gauge transformations. 
The wave functions are invariant with respect
to these transformations up to certain phase factors \cite{DJT,ja1},
 \be
\label{bc}
\Psi(\vecb{x}+ \vecb{a}, \vecb{y}) &=& e^{-2\pi i k \vecb{a} \vecb{y}} 
\Psi(\vecb{x},\vecb{y}) \, , \nonumber \\
\Psi(\vecb{x},\vecb{y}+\vecb{a}) &=& e^{2\pi i k \vecb{a} \vecb{x} } \Psi(\vecb{x},\vecb{y}) \, ,
  \ee
 where $\vecb{x} = \vecb{C}_1 L/(4\pi),  \vecb{y} = \vecb{C}_2 L/ (4\pi) $. It is enough then to consider
the motion over the (miltidimensional) {\it dual torus} $T_G \times T_G$ with the maximal torus $T_G$ representing the elementary cell of the coroot lattice. 
For $SU(2)$ when $r=1$, the dual torus corresponds to $x,y \in [0,1]$.  

 The presence of the phase factors \p{bc} represents a certain complication 
compared to the $4d$ case where such factors are absent. 
The effective BO Hamiltonian is also somewhat more complicated
than in the nonchiral $4d$ theories (where it is just a free Laplacian). 
Here it represents a multidimensional generalization
of the Landau--Dubrovin--Krichever--Novikov  Hamiltonian describing the motion 
in a planar magnetic field 
\cite{Novikov1,Novikov2}. For the group of rank $r$, the effective Hamiltonian reads 
 \be
    \label{HeffN}
    H \ =\ \frac {g^2}{2L^2} \left[  {({P}_{j}^a + {\cal A}_{j}^a)^2} +  {\cal B}^{ab}
 (\psi^a \bar\psi^b - \bar\psi^b \psi^a) \right] 
    \ee
with the matrix-valued $B^{ab} = \epsilon_{ij} \partial^a_i {\cal A}_j^b $ [and  the effective vector potentials 
${\cal A}_{j}^a(\vecb{C}_j)$ having nothing
to do,  of course, with the gauge fields of the original theory \p{LN1}]

But the most  serious $3d$ complication is that it {\it is} 
   not enough here to analyze the effective Hamiltonian 
to the leading BO order, but one-loop corrections should also be
taken into account. At the tree level, the magnetic field is homogeneous, 
       \be
     \label{postojan}
     {{\cal A}}_j^a &=&   -\frac {\kappa L^2}2 \epsilon_{jk} {C}_k^a \ , \nonumber \\
     {\cal B}^{ab} &=& \kappa L^2 \delta^{ab} \ ,
    \ee
 The loops bring about corrections that are singular at the ``corners'' 
of the  dual torus with $\vecb{C}_1$ and $\vecb{C}_2$ coinciding with 
the nodes of the lattice generated by the fundamental coweights. 
The number of such nodes in the dual torus is equal to the square of the order
of the center of the group. We will illustrate 
and explain these assertions later.

For $r=1$, the matrix-valued magnetic field becomes an 
ordinary one and the extra singular
loop-induced fields represent thin 
vortices 
\footnote{Their width is of order of mass $\sim g^2$, which is much less 
than the size of the dual torus $ 4\pi/L$.}
  placed in each of the $2 \cdot 2 = 4$  dual torus corners, 
$x,y = 0, 1/2$. For gluon loops, these lines carry the flux +1. In the limit 
$g^2L \to 0$, 
they become infinitely narrow
 Dirac strings. For fermion loops, the flux of each line is -1/2.
Half-integer magnetic fluxes are not admissible, they are not compatible with 
supersymmetry \cite{flux}. 
This refers, however, only to {\it net} fluxes and in our case 
the net magnetic flux of four lines with fluxes -1/2 each is  quite integer, 
$\Phi^{\rm ferm}_{\rm net} = -2$.       

An accurate analysis of Refs.\cite{ja2,ja3} (see also Sect. 4.4) 
displays that the integer fluxes 
are irrelevant for the vacuum counting 
(Dirac strings are not observable) and the result \p{Ik2} for 
the index is obtained from the tree-level result $I^{\rm tree} (k,2) = k+1$ 
by the substitution $k \to k-1$
rather than $k \to k+1$ as one should have written if gluon loops were taken into account.

Gluon loops should be irrelevant for any group --- 
this is the only way to reproduce the result \p{IkN} and the general recipe \p{recipe}. 
However, an explicit and 
rigourous demonstration of this fact was lacking up to now. 
This paper is written to fill out this gap.
 
\section{ $\theta$ functions.}
\setcounter{equation}0

We review here certain mathematical facts 
concerning the properties of $\theta$ functions living on the coroot 
lattice of a Lie group that we will use 
in the following. We have no doubt that they are known to mathematicians, though we were not able to find a 
manual with their clear exposition. 
 
 To begin with, let us remind some properties of the ordinary $\theta$ functions \cite{Mumford}. 
They are analytic functions on a torus  playing the 
same role there as the ordinary polynomials do for the Riemann sphere.
The polynomials have a pole at infinity and  $\theta$ functions satisfy certain nontrivial  
quasiperiodic boundary conditions with respect to shifts
 along the cycles of the torus.  
A generic torus is characterized by a complex modular parameter $\tau$, but we will stick to the 
simplest choice $\tau = i$ so that the torus represents a square $x,y \in [0,1]$ ( $z = x+iy$) glued around.

The simplest basic $\theta$-function satisfies the boundary conditions
 \be
\label{thet1bc}
 \theta(z+1) &=& \theta(z) \, ,\nonumber \\
 \theta(z+i) &=& e^{ \pi (1 - 2iz)} \theta(z) \, .
 \ee
This defines a {\it unique} (up to a constant complex factor) analytic function. Its explicit
form is
 \be
\label{thet1}
\theta(z) \ =\ \sum_{n = -\infty}^\infty \, \exp\{- \pi n^2 + 2\pi i n z \} \, .
 \ee
This function (call it theta function of level 1 and introduce an alternative
notation $\theta(z) \equiv Q^1(z)$) has only one zero in the square $x,y \in [0,1]$ --- 
right in its middle, 
$\theta(\frac {1+i}2 ) = 0$.

It will also be convenient for us in the following to use the function
 \be
\lb{pi}
\pi(z) \ =\ \frac {\theta \left( z - \frac {1+i}2 \right)}
{\theta'\left(\frac {1+i}2 \right)} \, .
 \ee
It has zeroes on the square lattice including the origin, where it behaves 
as $\pi(z) = z + O(z^2)$. It satisfies 
the boundary conditions 
  \be
\label{pibc}
 \pi(z+1) &=& \pi(z) \, ,\nonumber \\
 \pi(z+i) &=& -e^{ - 2i \pi z} \pi(z) \, .
 \ee
The function $\pi(z)$ is expressed into the function $\sigma(z)$
 defined in 
Eq.(8.171) of Ref.\cite{Ryzhik} with the choice 
  $\omega_1 = 1/2, \omega_2 = i/2$ for the half-periods as
 \be
\lb{pisigma}
 \pi(z) =  \exp\left\{ - \zeta\left(\frac 12 \right) z^2 + i \pi z \right\}\,
 \sigma(z)\, ,
 \ee
[the function $\zeta(z)$ to be defined later in \p{zeta}].

For any integer $q > 0$, one can define theta functions
of level $q$ satisfying
 \be
\label{thetqbc}
 Q^q(z+1) &=& Q^q(z) \, ,\nonumber \\
 Q^q(z+i) &=& e^{q \pi  (1 - 2iz)} Q^q(z) \, .
 \ee
A product of two such functions of levels $q$ and $q'$ gives a function of level $q+q'$.  

The functions satisfying (\ref{thetqbc}) lie in  vector space of dimension $q$. The basis
in this vector space can be chosen as 
 \be
\label{Qqm}
Q^q_m(z) \ \equiv \  \theta_{m/q,0} (qz, iq)\ ({\rm Mumford's\  notation}) \nn  
\ =\ \sum_{n = -\infty}^\infty \, \exp\left\{- \pi q \left(n + \frac mq \right)^2 + 2\pi i q z 
\left( n + \frac mq \right)\right\} \, , \nn
 m = 0, \ldots, q-1 \, .
 \ee

 Generically, a function $Q^q(z)$ 
has  $q$ simple zeros. A particular function of level 4,  
 \be
\lb{Pidef}
\Pi(z) = \ Q^4_3(z)- Q^4_1(z) \ =\ C e^{-2\pi i z} \, \pi (2z) 
 \ee
 will play a special role in our discussion. 
\footnote{The function \p{Pidef} and its analogs for higher groups to be discussed later 
were introduced in \cite{Pi,Eli,Laba} where the spectrum of pure CS theory was studied.}

$\Pi(z)$ is odd in $z$ and has four 
zeros at the corners $z = 0, \frac 12, \frac i2, \frac {1+i}2$. 
Notice that on top of \p{thetqbc} it also satisfies certain quasiperiodicity conditions 
with respect to half-integer shifts,
 \be
\lb{halfshift}
 \Pi(z+ 1/2) &=& -\Pi(z) \nn
\Pi(z + i/2) &=& -e^{\pi - 4\pi i z} \Pi(z) \, .
 \ee 

The vacuum wave functions of the theory \p{LN1} with $SU(2)$ gauge group are expressed into $\theta$ 
functions \p{Qqm}. For a group of rank $r >1$, we need $\theta$ 
functions of $r$ complex variables satisfying certain quasiperiodic boundary conditions on the 
coroot lattice of the corresponding group. To treat them, we will choose an inductive pragmatic approach
listing only some necessary (and sufficient for us) facts and discussing first the simplest $SU(3)$ 
case and generalizing to other groups afterwards.

\subsection{ $SU(3)$.}
 
 \begin{figure}[t]
\begin{center}
\includegraphics[width=3in]{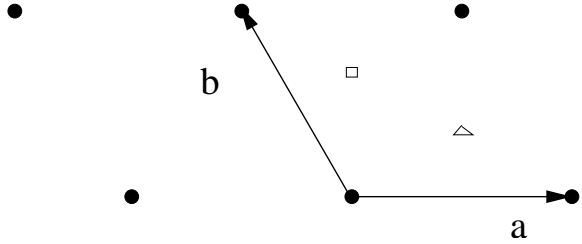}
\end{center}
\caption{Coroot lattice for $SU(3)$. The points marked by $\Box$ and 
$\triangle$ are fundamental coweights.}
\label{SU3lattice}
\end{figure}

Let $h = \ z^3 \lambda^3 + z^8 \lambda^8$ 
be  an element of the Cartan subalgebra of the complexified $su(3)$ algebra. 
The coroot lattice is depicted in Fig. \ref{SU3lattice}. 
We will be interested in the functions 
$\theta(\vecb{z})$ satisfying the following quasiperiodicity conditions
  \be
\lb{bcSU3}
\theta(\vecb{z} + \vecb{a}) \ =\ \theta(\vecb{z} + \vecb{b}) \ =\ \theta(\vecb{z}) \, , \nn
\theta(\vecb{z} + i\vecb{a}) \ =\ \exp \{k[2\pi - 4\pi i \vecb{z}\vecb{a}]\}   \theta(\vecb{z}) \, , \nn
\theta(\vecb{z} + i\vecb{b}) \ =\ \exp \{k[2\pi - 4\pi i \vecb{z}\vecb{b}]\}   \theta(\vecb{z}) \, , 
   \ee
where 
  \be
  \lb{ab}
\vecb{a} = (1,0), \ \ \ \ \ \ \ \ \ \ \ \ \vecb{b} = (-1/2,\sqrt{3}/2)
  \ee
 are the simple coroots. It follows that 
$$\theta(\vecb{z} + i\vecb{a}+ i\vecb{b} ) \ =\ \exp 
\{k[2\pi - 4\pi i \vecb{z}(\vecb{a}+\vecb{b})]\}   \theta(\vecb{z}) . $$ 
The property $\exp\{2\pi i \vecb{a} \vecg{\lambda} \} =  
\exp\{2\pi i \vecb{b} \vecg{\lambda} \} = 1$ holds. With the  metric choice 
  \be
\lb{nashKilling}
\langle h, g \rangle = \frac 12 {\rm Tr} \{ hg \} = \vecb{h} \vecb{g} \, ,
 \ee 
the simple coroots have the length 1. 

Speaking of the {\it roots} (defined according to $[h, E_\alpha] = \alpha(h)E_\alpha$ 
for positive root vectors $E_\alpha$), 
they represent the linear forms 
  \be
\lb{roots}
\alpha_a(\vecb{z}) = 2 \vecb{a} \cdot \vecb{z} \equiv 2 z^{(a)} , \ \ \ \ \  
\alpha_b(\vecb{z}) = 2 \vecb{b} \cdot \vecb{z} \equiv 2 z^{(b)}, \nn
  \alpha_{a+b}(\vecb{z}) = 
2 (\vecb{a}+ \vecb{b}) \cdot \vecb{z} \equiv 2 z^{(a+b)}
  \ee
   such that $\alpha_a(\vecb{a}) = \alpha_b(\vecb{b}) = 
\alpha_{a+b}(\vecb{a} + \vecb{b}) \ = 2$.

For a given
integer level $k$, the functions satisfying \p{bcSU3} form the vector space of dimension $3k^2$.
The product of two functions of levels $k,k'$ gives a function of level $k+k'$.

Three basis functions of level $k=1$ can be chosen in the form 
 \be
  \label{psi3k1}
  \Psi_0 &=& \sum_{\vecb{n}} \exp \left\{ -2\pi \vecb{n}^2 
+ 4\pi i \vecb{z} \vecb{n} \right \} \ , \nonumber   \\
  \Psi_\triangle &=& \sum_{\vecb{n}} \exp \left\{ -2\pi (\vecb{n} + 
  \triangle\!\!\!\!\!\triangle)^2 
   + 4\pi i \vecb{z} (\vecb{n} + \triangle\!\!\!\!\!\triangle) \right \} \ , 
\nonumber \\                    
      \Psi_\Box &=& \sum_{\vecb{n}} \exp \left\{ -2\pi (\vecb{n} + 
  \Box\!\!\!\!\!\Box)^2 
   + 4\pi i \vecb{z} (\vecb{n} + \Box\!\!\!\!\!\Box) \right \}\ ,           
   \ee
where the sum runs over the nodes of the lattice and 
\be
\lb{fundcoweight}
 \triangle\!\!\!\!\!\triangle = 
(2\vecb{a} + \vecb{b})/3,\ \  \Box\!\!\!\!\!\Box = (\vecb{a} + 2\vecb{b})/3
 \ee
  are the fundamental coweights.
\footnote{A fundamental coweight is an element of the Cartan 
subalgebra with zero projections
on all simple coroots but one, the nonzero projection being equal to $1/2$ 
in our normalization,
$\triangle\!\!\!\!\!\triangle \vecb{a} = \Box\!\!\!\!\!\Box \vecb{b} = 1/2$. 
Note that $\exp\{2\pi i \triangle^a \lambda^a\} = 
{\rm diag} (e^{-2\pi i/3},e^{-2\pi i/3}, e^{-2\pi i/3})   $ and   
$\exp\{2\pi i \Box^a \lambda^a\} = 
{\rm diag} (e^{2\pi i/3},e^{2\pi i/3}, e^{2\pi i/3})$ --- the elements of the 
center of $SU(3)$ (as was mentioned, for the coroots, 
these exponentials give the unit matrix).   }

There are two functions of this kind representing a particular interest.
One of them is expressed as
 \be
  \lb{TetSU3}
\Theta^{SU(3)}(\vecb{z}) \ =\ \Pi\left(\frac {z^8}{\sqrt{3}} \right) 
\Pi\left( \frac {z^3}2 + \frac {z^8}{2 \sqrt{3}}  
\right) \, \Pi \left( \frac {z^3}2 - \frac {z^8}{2 \sqrt{3}}  
\right) \nn
 = \ C\exp\left\{-2\pi i \left( z^3 + \frac {z^8}{\sqrt{3}} \right) \right\}
\pi\left( \frac {2 z^8}{\sqrt{3}} \right) 
\pi\left( z^3 + \frac { z^8}{\sqrt{3}} \right) 
\pi\left( z^3 - \frac { z^8}{\sqrt{3}} \right) 
 \ee
with $\Pi(z)$ defined in \p{Pidef}
[the arguments of the three functions $\pi$ in the RHS of Eq.\p{TetSU3} being
the fundamental weights ].  
It satisfies the boundary conditions 
\p{bcSU3} with $k=1$ and represents thus
a certain linear combination of the functions \p{psi3k1}

In contrast to the functions \p{psi3k1}, the function \p{TetSU3} has a simple structure of zeroes. 
It has the zeroes of the 3-d order in the nodes of the lattice in 
Fig. \ref{SU3lattice} and simple zeroes
on its edges, 
$ \frac {2z^8} { \sqrt{3} } \ = p+iq $ or 
 $z^3 + \frac {z^8}{ \sqrt{3}} =   p+iq $
or  $z^3 - \frac {z^8}{ \sqrt{3}} =  
 p+iq$.

 Consider now the function 
\be
\lb{PiSU3}
\Pi^{SU(3)}(\vecb{z}) \ =\ \Pi(z^3) \Pi\left( \frac {-z^3 + z^8 \sqrt{3} }2 
\right) \, \Pi\left( \frac {z^3 + z^8 \sqrt{3} }2 
\right)
 \ee
In this case, the arguments of the three functions $\Pi$  are  
the 
positive roots \p{roots} with the factor 2 removed.  
One can observe that the function $\Pi^{SU(3)}(\vecb{z})$
 satisfies the boundary conditions \p{bcSU3} with $k=3$. 

The structure of the zeroes of the function \p{PiSU3} 
is similar for that of the function \p{TetSU3}, but is more dense. It has zeroes of the third order
at the nodes of the coroot lattice and also at the points  
   \be
\lb{kovesa}
\left\{  \begin{array}{c} 
{\rm Re} (\vecb{z}) = \vecb{0}, 
\triangle\!\!\!\!\!\triangle, \Box\!\!\!\!\!\Box  \\
 {\rm Im} (\vecb{z}) = \vecb{0}, 
\triangle\!\!\!\!\!\triangle, \Box\!\!\!\!\!\Box
  \end{array} \right.  \, ,
\ee
 the fundamental coweights 
$\triangle\!\!\!\!\!\triangle, \Box\!\!\!\!\!\Box$ 
being defined in \p{fundcoweight}.
There are altogether 9 such points in $T_{SU(3)} \times T_{SU(3)}$.

\subsection{$SU(N)$, $Sp(4)$, $G_2$, etc.}

These definitions and observations can be generalized for any group. 

{\it (i)} $SU(N)$. Consider the  coroot lattice of $SU(N)$ generated
by its $N-1$ simple coroots $\vecb{a}^s$. Consider the functions $\theta (\vecb{z})$ that are periodic under the shifts 
$ \vecb{z} \to \vecb{z} + \vecb{a}^s $ and satisfy the conditions
 \be
\lb{bcSUN}
\theta (\vecb{z} + i \vecb{a}^s) \ =\ \exp \{k[2\pi - 4\pi i \vecb{z} \vecb{a}^s]\}   \theta (\vecb{z}) \, .
 \ee
The functions satisfying \p{bcSUN} form the vector space of dimension 
$Nk^{N-1}$. 

Consider the function 
\be
\lb{Piprodroots}
\Pi^{SU(N)}(\vecb{z}) \ =\ \prod_p \Pi\left[ \frac{\alpha_p(\vecb{z})}2 \right] 
 \ee
where the product runs over all $N(N-1)/2$ positive roots. 

One can be convinced that this function satisfies the boundary 
conditions \p{bcSUN} with $k = c_V = N$.
Indeed, consider the shift $\vecb{z} \to \vecb{z} + i\vecb{a}$ for a 
some particular coroot $\vecb{a}$. The argument in one of the factors in \p{Piprodroots} [with 
$\alpha_a(\vecb{z})$] is shifted by $i$. There are also $2(N-2)$ positive roots $\alpha_{\tilde p}$ 
with $\alpha_{\tilde p} (\vecb{a}) = -1$. Thus, the argument of $2(N-2)$ factors in \p{Piprodroots}
is shifted by $-i/2$. We can use now the boundary conditions \p{thetqbc} and \p{halfshift}.
Let us concentrate on the $\vecb{z}$ -- independent term  in the exponential. It is equal to 
 \be
\lb{phaseSUN}
 4\pi + 2(N-2) \pi \ =\ 2N\pi \, ,
  \ee
which matches \p{bcSUN} with $k=N$. The linear in $\vecb{z}$ terms also match.

The identity \p{phaseSUN} is actually a manifestation of the  general identity
    \footnote{ Though it is known to experts \cite{Kacpriv}, we were
 not able to find it in the standard textbooks. Its elementary proof is 
outlined in Appendix.}

   \be
 \lb{sumkvadratov}
\sum_{p={\rm positive\  roots}} [\alpha_p(\vecb{h}) ]^2 \ =\ 2 {c_V}  
\vecb{h}^2 
  \ee
valid not only for coroots $\vecb{a}$ in $SU(N)$, but for  any element $\vecb{h}$  
in the Cartan subalgebra of any Lie algebra.

There are $N^2$ special points where the function \p{Piprodroots} has zeroes
of order $N(N-1)/2$: Re($\vecb{z}$) and  Im($\vecb{z}$) 
can be zero or coincide with one of $N-1$ fundamental coweights of $SU(N)$.  
  
\begin{figure}[t]
\begin{center}
\includegraphics[width=3in]{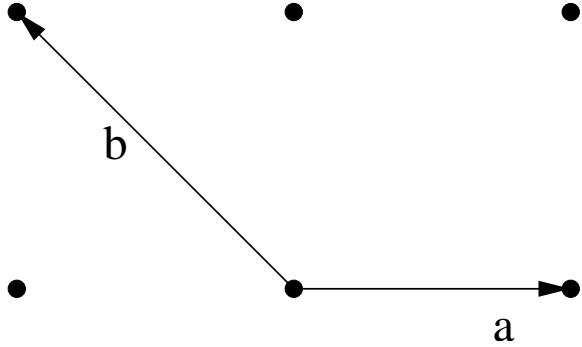}
\end{center}
\caption{Coroot lattice for $Sp(4)$.}
\label{Sp4lattice}
\end{figure}

{\it (ii)} $Sp(4)$. It is convenient to choose the following orthogonal basis
in the Cartan subalgebra, 
$e_1 = {\rm diag} (0,1,-1,0)$, \  $e_2 = {\rm diag} (1,0,0,-1)$. In this
basis, the simple coroots $\alpha^\vee = e_1, \beta^\vee = e_2-e_1$ are represented 
by the vectors  $\vecb{a} = \{1,0\}$ and 
$\vecb{b} = \{-1,1\}$. These simple coroots generate 
the coroot lattice depicted in Fig.\ref{Sp4lattice}. 
The maximal torus represents just a square [not 
a rhombus as for $SU(3)$]. There are two short 
($\vecb{a}, \vecb{b} + \vecb{a}$) and two long 
($\vecb{b}, \vecb{b} + 2\vecb{a}$) positive coroots.

The relevant $\theta$ functions satisfy the boundary conditions
  \be
\lb{bcSp4}
\theta(\vecb{z} + i\vecb{a}) \ =\ \exp \{k[2\pi - 4\pi i \vecb{z}\vecb{a}]\}   \theta(\vecb{z}) \, , \nn
\theta(\vecb{z} + i\vecb{b}) \ =\ \exp \{k[4\pi - 4\pi i \vecb{z}\vecb{b}]\}   \theta(\vecb{z}) \, .
 \ee 
with an integer $k$. They form the vector space of dimension
$4k^2$. The appearance of $e^{4\pi k}$ rather 
than $e^{2\pi k}$ in the exponential factor in the second line
is due to the fact that $\vecb{b}$ is a long coroot, 
$\vecb{b}^2 = 2$, while $\vecb{a}^2 = 1$.

The conditions \p{bcSp4} imply that 
 \be
\lb{bcany}
 \theta(\vecb{z} + i\vecb{c}) \ =\ \exp \{k[2\pi \vecb{c}^2 - 4\pi i \vecb{z}\vecb{c}]\}   \theta(\vecb{z}) 
 \ee
for all other vectors $\vecb{c}$ in the coroot lattice. 

Consider the function
\be
\lb{PiSp4}
\Pi^{Sp(4)}(\vecb{z}) \ =\ \Pi(z_1) \Pi(z_2) 
\Pi\left( \frac {z_1 + z_2 }2 \right)  \Pi\left( \frac {z_1 - z_2 }2 \right)\, .
 \ee
As its analogs written before, the function \p{PiSp4} 
is represented in the form \p{Piprodroots}. It satisfies the boundary 
conditions \p{bcSp4} with $k = c_V[Sp(4)] = 3$, which follows again from  \p{sumkvadratov}.

The function \p{PiSp4} has  zeroes of the 4-th order at four special points
on $T_{Sp(4)} \times T_{Sp(4)}$:\ 
${\rm Re}(z_1)  = {\rm Re} (z_2) = 0, 1/2$ and $ {\rm Im} (z_1) = {\rm Im} (z_2) = 0, 1/2$, 
the  point $z_1  = z_2 = 1/2$ 
 corresponding to the nontrivial element 
$-1\!\!\!\!1$ of the center $Z_2$ of $Sp(4)$.

Besides, it has zeroes of the second order when only $\Pi(z_1)$ and $\Pi(z_2)$
vanish. There are 12 such zeroes at
 \be
\lb{12zeroes}
\left\{    \begin{array}{c}   z_1 = 0  \\ z_2 = \frac 12, \frac i2 ,  
\frac {1+i}2 \end{array} 
\right. \, , \ \ \ \ \ \ \ \ \left\{    \begin{array}{c}   z_1 = \frac 12
 \\ z_2 = 0, \frac i2 , \frac {1+i}2 \end{array} \right.\, , \nn
\left\{    \begin{array}{c}   z_1 = \frac i2  \\ z_2 = 0, \frac 12, 
\frac {1+i}2  \end{array} 
\right. \, , \ \ \ \ \ \ \ \ \left\{    \begin{array}{c}   z_1 = \frac {1+i}2
 \\ z_2 = 0, \frac 12,  \frac i2  
 \end{array} \right.\,
  \ee

{\it (iii)} $G_2$. The coroot lattice has the 
same hexagonal form as for $SU(3)$, but it is generated
now by the simple coroots $\vecb{a}$ and $\vecb{b} = (-3/2, \sqrt{3}/2)$ . 
There are 6 positive coroots:
3 short coroots $\vecb{a}, \vecb{b} + \vecb{a}, \vecb{b} + 2\vecb{a}$ and 
3 long coroots
$\vecb{b},   \vecb{b} + 3\vecb{a}, 2\vecb{b} + 3\vecb{a}$. 
The $\theta$ functions satisfying 
\p{bcSU3} satisfy
also the property
  \be
\lb{bcG2}
\theta(\vecb{z} + i\vecb{b}) \ =\ \exp \{k[6\pi - 4\pi i \vecb{z}\vecb{b}]\}   \theta(\vecb{z}) \, ,
 \ee  
the norm $\vecb{b}^2 = 3$ giving the  term 
$2\pi k \vecb{b}^2 = 6\pi k$ in the exponent.  

 \begin{figure}[t]
\begin{center}
\includegraphics[width=3in]{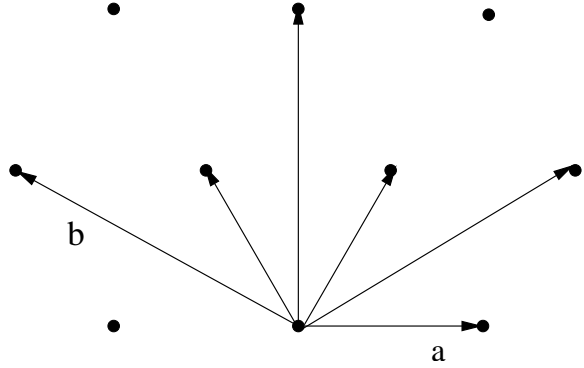}
\end{center}
\caption{Positive coroots for $G_2$.}
\label{G2lattice}
\end{figure}

In virtue of \p{sumkvadratov}, the function
\be
\lb{PiG2}
\Pi^{G_2}(\vecb{z}) \ =\ \Pi(z^3) \Pi\left( \frac {-z^3 + z^8 \sqrt{3} }2 
\right) \, \Pi\left( \frac {z^3 + z^8 \sqrt{3} }2 \right)  
\times \nn
\Pi\left( \frac{z^8}{\sqrt{3}} \right)
\Pi\left( \frac {z^3 \sqrt{3} + z^8}{2 \sqrt{3} } 
\right) \Pi\left( \frac {z^3 \sqrt{3} - z^8}{2 \sqrt{3} } 
\right) 
 \ee 
(we kept the $SU(3)$ notation for the components of $\vecb{z}$) satisfies the boundary 
conditions \p{bcSU3} (and its colloraries \p{bcany} 
for all other vectors in the coroot lattice) with $k = c_V[G_2] = 4$.

There is only one special point on $T_{G_2} \times T_{G_2}$ ($\vecb{z} = 0$)
 where the function \p{PiG2} has a zero of 6-th order.
This conforms to the fact that $G_2$ has no centre.

Then there are 8 zeroes of the 3-d order at the points 
 \be
\lb{8zeroesG2}
\left\{ \begin{array}{c}
{\rm Re} (\vec{z}) = 0, \triangle\!\!\!\!\!\triangle, \Box\!\!\!\!\!\Box \, , 
\\
{\rm Im}(\vec{z}) = \triangle\!\!\!\!\!\triangle, \Box\!\!\!\!\!\Box
\end{array} \right. \, , \ \ \ \ \ \ \ \ \ 
\left\{ \begin{array}{c}
{\rm Re} (\vec{z}) =  \triangle\!\!\!\!\!\triangle, \Box\!\!\!\!\!\Box \, , 
\\
{\rm Im}(\vec{z}) = 0
\end{array} \right. \, , 
   \ee
where only the factors 
$\Pi(z^3)$ and $\Pi\left( \frac {\pm z^3 + z^8 \sqrt{3}}2 \right)$ corresponding to 
the long roots of $G_2$ vanish. 

Finally, there are the zeroes of the 2-nd order when only 
two factors in \p{PiG2} corresponding to a pair of orthogonal long and 
short roots, like $\Pi(z^3)$ and $\Pi\left( \frac {z^8}{\sqrt{3}} \right)$,  vanish.
There are 9 such points,
\be
\lb{9zeroesG2}
\left\{    \begin{array}{c}   z^3 = 0  \\ z^8 = \frac {\sqrt{3}} 2, \frac {i\sqrt{3}} 2, \frac {(1+i)\sqrt{3}} 2, \end{array} 
\right.  , \  
\left\{    \begin{array}{c}   z^3 = \pm \frac 14  \\ z^8 = \frac {\sqrt{3}} 4 \end{array} 
\right. , \   
\left\{    \begin{array}{c}   z^3 = \pm \frac {i} 4  \\ z^8 = \frac {i\sqrt{3}} 4 \end{array} 
\right. , \  
\left\{    \begin{array}{c}   z^3 = \pm \frac {(1+i)}4  \\ z^8 = \frac {(1+i)\sqrt{3}} 4 \end{array} 
\right.
 \ee

Generalization of these constructions to all other groups is straightforward.

\section{Dirac strings and multidimensional Dirac strings.}

\setcounter{equation}0

\subsection{Effective theory}

The effective wave functions depend on $r$ slow complex bosonic variables $z^a = x^a + iy^a$, 
their conjugates, and their holomorphic fermionic superpartners $\psi^a$. The effective theory belongs
to the class of complex supersymmetric sigma models introduced in \cite{Hull} and studied in details 
in \cite{DiracSQM}. In our case, the metric is flat and the superfield action reads
 \be
\lb{actWZ} 
 S \ =\ \int dt d^2\theta \left[ - \frac 14 D Z^a \bar D \bar Z^a + W(Z^a, \bar Z^a) \right] \, ,
 \ee
where 
$$ D \ =\ \frac \partial {\partial \theta} - i \bar \theta \frac 
{ \partial} {\partial t}, \ \ 
  \bar D \ =\ -\frac \partial {\partial \bar \theta} + i 
 \theta \frac { \partial} {\partial t} $$
are the supersymmetric covariant derivatives and $Z^a = z^a + \sqrt{2} \theta \psi^a - i \theta \bar \theta
\dot{z}^a $ depending on $t_L = t - i \theta \bar \theta$  and  
$\bar Z^a = \bar z^a - \sqrt{2} \bar\theta 
\bar\psi^a  + i \theta \bar \theta
\dot{\bar z}^a $ depending on $t_R = t + i \theta \bar \theta$ are chiral superfields, 
$\bar D Z^a = D \bar Z^a = 0$. The particular form of the real prepotential  $W(Z^a, \bar Z^a)$
will be shortly revealed.

The nilpotent N\"other supercharges derived from \p{actWZ} are 
 \be
\lb{QQbar} 
Q = \sqrt{2} (P_a + i \partial_a W) \psi^a, \ \ \ 
\bar Q = \sqrt{2} (\bar P_a - i \bar \partial_a W) \bar \psi^a  \, ,
\ee
 where $\partial_a = (\partial_1^a - i\partial_2^a)/2 $ and 
$ \bar \partial_a = 
(\partial_1^a + i\partial_2^a)/2$ are holomorphic and antiholomorphic derivatives. 
The Hamiltonian is 
 \be
\lb{HeffWpsi} 
 H \ =\ (\bar P_a - i \bar \partial_a W) ( P_a + i  \partial_a W) -    
2(\partial_a \bar \partial_b W)
\psi^a \bar \psi^b  \ .
 \ee
Being multiplied by a proper constant, it can be expressed in the form \p{HeffN}. Note, however,
that the effective vector potentials ${\cal A}_j^a$ entering   \p{HeffN} are not arbitrary,
 but are derived from a single real prepotential, 
${\cal A}_j^a \propto \epsilon_{jk} \partial^a_k W$. At the tree level,  
${\cal A}_j^a$ have the form \p{postojan}. This corresponds to 
$W^{\rm tree} = -\pi k \bar z^a z^a$. Loop corrections bring about extra 
effective gauge fields. 
Let us discuss their structure 
first in  the simple  $SU(2)$  ($r=1$) case and then for the 
groups of higher rank.

\subsection{$SU(2)$.}

As was mentioned before, gluon and fermion loops bring about thin vortices of fluxes 
+1 and $-1/2$, respectively. In our problem, there are only two spatial dimensions, 
but one can imagine the existence of the third orthogonal direction where  
 the vortices (representing now fluxes lines) extend. For the flux +1, 
the physics of such a line is the same as for a {\it Dirac string}, and that is 
how we will call it, also in two dimensions.

The Dirac string piercing the origin $z=0$ corresponds to 
 \be
\lb{Wstring}
W^{\rm string} \ =\ - \frac 12 \ln(z \bar z)  
 \ee
such that the holomorphic  potentials are 
 \be
\lb{Avorthol}
 {\cal A} = \frac {{\cal A}_1 -i{\cal A}_2}2 = i \partial W = - \frac i{2z} , \ \ \ \ \ \ \ 
\bar {\cal A} = - i \bar \partial W =  \frac i{2 \bar z}
\, .
\ee
and
\be
\lb{Avort}
 {\cal A}_j \ = \ -\frac{\epsilon_{jk} x_k}{\vecb{x}^2} \, .
 \ee 
The supercharges are
 \be
\lb{QQbarz}
 Q \ =\ -i\sqrt{2} \left( \frac{\partial}{\partial z} + \frac 1{2z} \right) 
\psi, \ \ \ \ \bar Q \ =\ -i\sqrt{2} \left( \frac{\partial}{\partial \bar z} -
 \frac 1{2\bar z} \right) \bar \psi \, .
 \ee
As was mentioned, the motion over the dual torus is finite. But let us forget it for a while
and consider the operators \p{QQbarz} acting on the wave functions that live on the infinite complex plane.
The spectrum of the corresponding Hamiltonian 
 \be
\lb{Hstring}  H\ =\  - \left(\bar \partial - \frac 1{2\bar z}   \right) 
\left( \partial + \frac 1{2 z}   \right) 
 \ee
(the Hamiltonian in the sectors $F=0$ and $F=1$ is the same) 
is then continuous. It is easy to see that the spectrum of \p{Hstring}
  coincides with the spectrum of the free 
Laplacian $-\bar  \partial \partial$: all the eigenstates of $H$ 
are obtained from 
the eigenstates of $-\bar  \partial \partial$ 
by multiplying the latter by the factor 
$\sqrt{\bar z/z } = e^{-i\phi}$.
\footnote{ We exclude from the spectrum the singular 
quasinormalizable zero energy  state 
 with the wave function 
\cite{Kiskis} 
 \be    
\lb{Kiskis}
 \Psi_0 \ = \ \frac 1{\sqrt{\bar z z} } \, .
 \ee}

Also for a finite motion, adding a Dirac string at some point does not affect the spectrum
of  the Hamiltonian and the wave functions are multiplied by the factor $e^{-i\phi}$. 
This means that an infinitely thin vortex of unit flux is actually not observable. Non-observability of  
 Dirac strings in the problem of motion of a scalar or a spinor particle 
in the field of a conventional 3-dimensional Dirac 
monopole with properly quantized charge is, of course, a well-known fact \cite{Wu}.

\subsection{$SU(3)$ and higher groups.}

\begin{figure}[t]
\begin{center}
\includegraphics[width=3in]{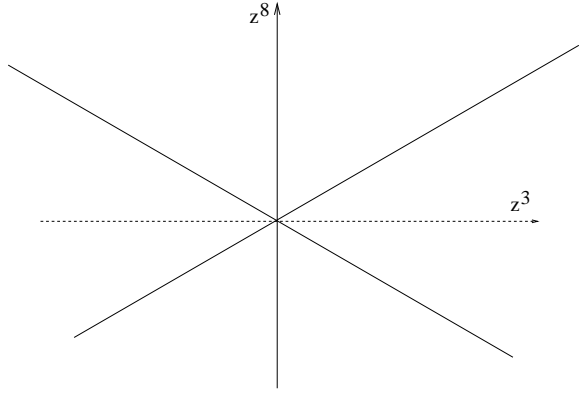}
\end{center}
\caption{Singularities of the potential for the $SU(3)$ string. 
Complex $z^a$ are represented by their real parts.}
\label{SU3string}
\end{figure}

For $SU(3)$, the index $a$ in \p{actWZ} takes two values, $a = 3,8$. Let us choose
\be
\lb{WSU3}
 W(z^a, \bar z^a) \ =\ -\frac 12 \left\{ \ln(\vecb{z} \vecb{a} ) + 
\ln(\vecb{z} \vecb{b} ) + \ln[\vecb{z} (\vecb{a} + \vecb{b}) ] \right \} \ +\ {\rm c.c.}
 \ee
with $\vecb{a}, \vecb{b}$ defined in \p{ab}. The  supercharges acquire the form 
 \be
\lb{Qaeff}
 Q = -i \sqrt{2} (\partial_a + i {\cal A}_a) \psi^a , \ \ \ \ \ \ 
\bar Q  = -i \sqrt{2} (\bar\partial_a + i \bar {\cal A}_a) \bar \psi^a
 \ee
with 
 \be
\lb{calAa}
{\cal A}_3 \ =\ \partial_3 W \ =\ -\frac i2 \left( \frac 1{z^3} + \frac 1{z^3 + z^8 \sqrt{3} } +
\frac 1{z^3 - z^8 \sqrt{3} } \right) \nn 
{\cal A}_8 \ =\ \partial_8 W \ =\ \frac {i\sqrt{3}}{2} \left( \frac 1{z^3 - z^8 \sqrt{3}} 
-   \frac 1{z^3 + z^8 \sqrt{3}} \right)
 \ee
The vector potentials \p{calAa} represent the $SU(3)$ counterpart of the standard Dirac string
\p{Avorthol}. They live on $\mathcal{C}^2$ and, in contrast to the usual Dirac string, are singular not just
at one point, but on 3 planes $z^3 = 0$ and $z^3 = \pm \sqrt{3} z^8$ (see Fig. \ref{SU3string}). 
As is clear, this object enjoys $O(2)$, but not $O(4)$ symmetry.

The supercharge $Q$ gives zero when acting on the function 
 \be
\lb{fSU3}
f_{SU(3)}(z^a, \bar z^a) \ =\ \sqrt{\frac {\bar z^3 [(\bar z^3)^2 - 3(\bar z^8)^2]}{ z^3 [( z^3)^2 - 3( z^8)^2]}}
\, .
 \ee 
The function \p{fSU3} is uniquely defined on $\mathcal{C}^2$. It is the $SU(3)$ counterpart of the factor 
$e^{-i\phi}$ for $SU(2)$. Note now that the spectrum of the 
 Hamiltonian with the $SU(3)$ Dirac string
added coincides with the spectrum of the Hamiltonian without such string. The wave functions of the former
are obtained from the wave functions of the latter by multiplication over the factor \p{fSU3}. The Dirac string
\p{calAa} is unobservable! 

A generalization for an arbitrary group is straighforward. We should consider instead of \p{WSU3} 
the function
\be
\lb{Wany}
 W(z^a, \bar z^a) \ =\ -\frac 12 \sum_p \ln[\alpha_p(\vecb{z})]  
 \ +\ {\rm c.c.} \, ,
 \ee
where the sum runs over all positive roots. Note that there are {\it three different} generalized Dirac
strings living on $\mathcal{C}^2$ corresponding to three different simple groups of rank 2,
 there are 3 different strings for  $\mathcal{C}^3$, etc. 
The   Hamiltonian involving 
an extra multidimensional Dirac string has the same spectrum as the Hamiltonian without such string, with
the wave functions being multiplied by the (uniquely defined on $\mathcal{C}^r$ ) factor 
 
\be
\lb{fany}
f_G(\vecb{z},  \vecb{\bar z}) \ =\ \sqrt{\prod_p \frac {\alpha_p( \vecb{\bar z})}{\alpha_p(\vecb{z}) } }\, .
 \ee 

The conventional 3-dimensional Dirac strings associated with the monopoles also have a multidimensional
generalization. Multidimensional analogs of monopoles were constructed in \cite{Blok}. They appear when 
constructing the effective Hamiltonian in the chiral (3+1)- supersymmetric gauge theories. This Hamiltonian
depends on $3r$ slow variables. The effective multidimensional vector potentials that live in 
$\mathcal{R}^{3r}$ 
are singular on hyperplanes 
of codimension 2 whose structure is similar to that displayed in 
Fig.\ref{SU3string}. The kinship of these two different
problems is natural: in three dimensions, mass fermion term breaks parity and hence the theory \p{LN1} {\it is} chiral.

\section{Index of the strings}
\setcounter{equation}0

\subsection{$SU(2)$}

As was mentioned, the string \p{Avorthol} carries the unit flux. 
To see this, one has to regularize it  replacing  \p{Avorthol} by
   \be
\lb{Avortreg}
 {\cal A} =  - \frac {i \bar z} {2(\bar z z + m^2)} , \ \ \ \ \ \ \ \bar {\cal A}  =  
\frac {i z} {2 (\bar z z + m^2)} \, .
\, .
\ee
The corresponding magnetic field is 
 \be
\lb{Bvortreg} 
 B(\bar z, z) = 2i\left( \bar \partial {\cal A} -
 \partial  \bar {\cal A} \right) 
\ =\  \frac{2m^2} {(\bar z z + m^2)^2} \, .
  \ee
Hence
 \be 
\lb{flux}
I \ =\ \frac \Phi {2\pi} \ =\ \frac 1 {2\pi} \int dz d\bar z \,  
B(\bar z, z) \ =\ 1 \,. 
  \ee
The integral is saturated by the region of small $|z| \sim m$. 

The famous
Atiyah-Singer theorem seems to dictate for the spectrum of the Dirac operator
(our supersymmetric problem is equivalent to the Dirac problem) 
in the field of unit flux to involve a zero mode. However, 
the best what we can get by solving the zero mode equation $Q \Psi = 0$ with 
the gauge field   \p{Avortreg} is the function
  \be
\lb{zeromode}
 \Psi_0 \ = \ \frac {F(\bar z)} {\sqrt{\bar z z + m^2} }
 \ee
with an arbitrary antiholomorphic $F(\bar z)$. This is not a  ``catholic'' zero mode because
 the normalization integral diverges
logarithmically or worse. If choosing $F(\bar z) = 1$ and lifting
the regularization, it goes to \p{Kiskis}.  

  In fact, the AS theorem 
applies only to {\it compact} 
manifolds where the spectrum is discrete. Thus, to get a nice 
normalizable zero mode, 
we need to compactify the complex plane. 
It is usually done by replacing
$\mathcal{C} \to S^2$ \cite{Wu} , but compactification on the torus is 
also possible \cite{Nambu}.

The latter implies nontrivial boundary conditions for the
wave functions. In the presence of the
 magnetic flux, $\Phi = 2\pi q$ 
with integer nonzero $q$, the wave functions 
are not just periodic, but involve certain phase factors, 
  \be
\lb{twist}
 \Psi(x+1,y) \ =\ e^{i\alpha(x,y)}  \Psi(x,y) \, , \nn
  \Psi(x,y+1) \ =\ e^{i\beta(x,y)}  \Psi(x,y)
 \ee
with the functions  $\alpha(x,y)$ and $\beta(x,y)$ satisfying the condition

 \be
\lb{obhod}
 \alpha(x,y) + \beta(x+1, y) - \alpha(x, y+1) - \beta(x, y) \ =\ 2\pi q \, .
 \ee
In the case under consideration, $q=1$. 
Different choices for the phases $\alpha,\beta$ are possible. One of the choices 
was presented in \p{bc} where one should 
replace $\vecb{x} \to x, \vecb{y} \to y, \vecb{a} \to 1$ and set $2k=q=1$. 
It is more convenient for us now to use an asymmetric 
choice $\alpha(x,y) = 0, \ \beta(x,y) = 2\pi x$. 
The toric zero mode should satisfy the boundary conditions \p{twist} and behave as \p{zeromode} near the origin. 
It is difficult to write an analytic expression for such a function in a generic massive case, 
but in the massless limit $m \to 0$, it can be easily done,
 \be
\lb{zeromodetorus}
 \Psi_0^{\rm torus} \ = \ 
\sqrt{ \frac {\bar \pi( z) } { \pi( z)  }} \, .
 \ee

For $ |z| \ll 1$, the wave function \p{zeromodetorus} behaves as 
$\sqrt{\bar z/ z} = e^{-i\phi}$.  

The function \p{zeromodetorus} satisfies the equation
    \be
\lb{eqzeromodetorus}
 (\partial + i  {\cal A} ) \Psi_0^{\rm torus} \ =\ 0
 \ee
with 
 \be
\lb{barA}
  {\cal A} \ =\ -\frac {i \pi'(z)}{2\pi(z)}
  \ee
If thinking in  terms of the whole complex plane ${\mathcal C}$, 
the vector potential in \p{barA} 
corresponds to a regular lattice of strings placed at $z = p + iq$ 
with integer $p,q$. 
Bearing in mind \p{pisigma}, it can be expressed as  
$$   {\cal A} =\  - \frac i2 \left[ \zeta(z) + i\pi  -2z \zeta (1/2) \right] ,$$
where $\zeta(z)$ is the Weierstrass
zeta function,
 \be
\lb{zeta}
\zeta(z) \ =\ \frac 1z + \sum_{\{pq\} \neq \{00\} } \left( \frac 1{z+p + iq} - \frac 1{p + iq} 
+ \frac z {(p + iq)^2}  \right)  
\nn \frac 1z - \int_0^z \left( {\cal P} (u) - \frac 1{u^2} \right) du \, .
 \ee

\subsection{Strings on $\mathcal{C}^r$ and the associated index integrals.}

{\it (a)} $SU(N)$.

For the effective Hamiltonian \p{HeffN}, the analog of \p{flux} reads \cite{ja1}
\be
\lb{fluxN}
I \ =\ \frac 1 {(2\pi)^r}  \int \prod_{a=1}^r  \prod_{j=1}^2 \, dC_j^a \, \det \| {\cal B}^{ab} \| \, ,
  \ee
where the integral is done over the relevant range of $C^a_j$. The result \p{fluxN} is obtained by replacing 
the functional integral for the index \p{Witind} by the ordinary one which is admissible in the semiclassical limit
$ \beta \to 0$ \cite{Cecotti}.   The determinant appears after integration over fermion variables.

For $SU(3)$,  the integral \p{fluxN}  was calculated in \cite{ja1}. For a (regularized) 
individual Dirac string \p{calAa} living on $\mathcal{C}^2$, 
the result  is $I=3$. Bearing in mind further generalizations 
for more complicated groups, let us describe this calculation in some more details
 choosing the simple coroot basis (cf. \cite{Blok}),
    \be
   \lb{simplezpsi}
   \psi^s =  \vecg{\psi} \vecb{a}^s, \ \ \ \ \  z^s =  \vecb{z} \vecb{a}^s \, .
    \ee

We  regularize the  superpotential \p{Wany} and write (bearing in mind that $z^p = \frac 12 
\alpha_p(\vecb{z})$ for
the simply laced groups)
   \be
\lb{Wanyreg}
 W \ =\ -\frac 12 \sum_p \ln \left[|z^p|^2 + m^2 \right]  
 \, ,
 \ee 
For $SU(3)$, the sum involves 3 terms.
The vector potentials in the root basis are 
 \be
\lb{ASU3prost}
{\cal A}^1 \equiv {\cal A}^{(a)} = {\cal A}_3 + \frac {1}{\sqrt{3}}
 {\cal A}_8  =  - \frac i2 \left(   \frac {\bar z^{(a)}}
{\bar z^{(a)}  z^{(a)} + m^2} +   \frac {\bar z^{(a+b)}} 
{\bar z^{(a+b)}  z^{(a+b)} + m^2}  \right) \, ,  \nn
  {\cal A}^2 \equiv {\cal A}^{(b)} = \frac {2}{\sqrt{3}} {\cal A}_8 =
 - \frac i2 \left(   \frac {\bar z^{(b)}}
{\bar z^{(b)}  z^{(b)} + m^2} +   \frac {\bar z^{(a+b)}} 
{\bar z^{(a+b)}  z^{(a+b)} + m^2}  \right) \, ,
 \ee
$z^{(a+b)} = z^{(a)} + z^{(b)}$.

They enter the effective supercharges expressed as
 \be
\lb{QSU3}
 Q = -i\sqrt{2} \psi^{(a)} \left[ \frac {\partial}{\partial z^{(a)}} + 
i {\cal A}^{(a)} \right] - i\sqrt{2} \psi^{(b)} \left[ \frac {\partial}{\partial z^{(b)}} + 
i {\cal A}^{(b)} \right]
 \ee

The magnetic fields ${\cal B}^{ss'} = 2i\left( \bar \partial^s {\cal A}^{s'} -
 \partial^{s'}  \bar {\cal A}^s \right) $ are 
\be
\lb{BvortSU3} 
{\cal B}^{11} = {\cal B}(\bar z^{(a)}, z^{(a)}) +  {\cal B}(\bar z^{(a+b)}, z^{(a+b)})\, , \nn
{\cal B}^{12} =  {\cal B}^{21} = {\cal B}(\bar z^{(a+b)}, z^{(a+b)}), \nn
 {\cal B}^{22} = {\cal B}(\bar z^{(b)}, z^{(b)}) +  {\cal B}(\bar z^{(a+b)}, z^{(a+b)}) 
\, ,
  \ee
where ${\cal B}(\bar z, z)$ is the universal function written in \p{Bvortreg}.

We obtain
 \be
 \lb{indSU3}
I \ =\ \frac 1{4\pi^2} \int \prod_{s=1,2} \  d\bar z^s d z^s \left( {\cal B}_a   {\cal B}_b + 
{\cal B}_a   {\cal B}_{a+b} + {\cal B}_b   {\cal B}_{a+b} \right)
 \ee
Each term in \p{indSU3} gives a unit contribution and we obtain the result $I = 3$. 

A similar calculation for $SU(4)$ gives the magnetic field matrix
  \be
\lb{BvortSU4} 
 \left( 
\begin{array}{ccc} {\cal B}_a +  {\cal B}_{a+b} + {\cal B}_{a+b+c} & {\cal B}_{a+b} + 
{\cal B}_{a+b+c} & {\cal B}_{a+b+c}\\
{\cal B}_{a+b} + {\cal B}_{a+b+c} & {\cal B}_{b} + {\cal B}_{a+b} + {\cal B}_{b+c} + {\cal B}_{a+b+c} &
{\cal B}_{b+c} + {\cal B}_{a+b+c} \\
 {\cal B}_{a+b+c} & {\cal B}_{b+c} + {\cal B}_{a+b+c} & {\cal B}_c + {\cal B}_{b+c} + {\cal B}_{a+b+c}
\end{array} \right) 
  \ee
in obvious notations ($a,b,c$ being the simple roots).
In the determinant, only the products of magnetic fields associated with all {\it different} 
positive roots survive, 
each such product giving a contribution $+1$ to the index integral. The number of such products is
 \be
\lb{Inumbers4}
I \ =\ \left \| \begin{array}{ccc} 3 & 2 & 1 \\ 2 & 4 & 2 \\ 1 & 2 & 3 \end{array} \right \| = 16 \, .
 \ee
For $SU(5)$ 
\footnote{To justify quite rigorously the estimates \p{Inumbers5} and \p{Inumbers6}, 
one has to demonstrate  that only the products of different positive roots survive in the determinant.
 To justify 
\p{InumbersN}, one has to prove in addition that the observed pattern for the determinants 
\p{Inumbers4} - \p{Inumbers6}
generalizes for any $N$. This is an interesting question to clarify.},
we obtain
    \be
\lb{Inumbers5}
I \ =\  \left \| \begin{array}{cccc} 4 & 3 & 2 & 1 \\ 3 & 6 & 4 & 2 \\ 2 & 4 & 6 & 3 \\ 
1& 2 & 3 & 4 \end{array} \right \| = 125 \, .
 \ee
The numbers on the diagonal of this matrix are the numbers of the positive roots 
involving a given simple root
$a$, $b$, $c$ or $d$ in their simple root 
expansion. The adjacent numbers  are the numbers of positive roots involving 
in the expansion {\it two}  roots $(a,b)$, $(b,c)$ and $(c,d)$.
For example, the positive roots involving both $b$ and $c$ are $b+c$, \ $a+b+c$, \ $b+c+d$,\ 
 $a+b+c+d$, and the 
corresponding matrix element is 4.  Next, 2 is the number of positive roots involving {\it three} roots $(a,b,c)$ and
$(b,c,d)$ in the expansion. 
Finally, there is only one root, $a+b+c+d$, involving all four simple roots 
in the expansion.

 For  $SU(6)$, the result is 
  \be
\lb{Inumbers6}
I \ =\  \left \| \begin{array}{ccccc} 5&4& 3 & 2 & 1 \\ 4 & 8 & 6 & 4 & 2  
\\ 3 & 6 & 9 & 6 & 3 \\
2 & 4 & 6 & 8 & 4 \\ 
1& 2 & 3 & 4 & 5 \end{array} \right \| = 1296 \, .
 \ee
For an arbitrary $N$, the index integral is (conjectured to be)
 \be
\lb{InumbersN}
 I \ =\ N^{N-2}
 \ee

  {\it b)} $Sp(4)$.

We keep the notations \p{simplezpsi} 
with the simple coroots $\vecb{b}, \vecb{a}$ as in Fig.2. There are altogether
two long coroots $\vecb{b}, \ \vecb{b} + 2\vecb{a}$ and two short ones  $\vecb{a}, \ \vecb{b} + \vecb{a}$.
Note that, while, for $SU(N)$, $z^p$ are all related to the roots as
  $z^p =  \frac 12 \alpha_p(\vecb{z})$, for $Sp(4)$, 
this is true only for the short coroots (and long roots) whereas for the long coroots (and short roots), 
$z^p$ and $\alpha_p(\vecb{z})$ just coincide. 

Choose the superpotential as in \p{Wanyreg}. The  supercharge has the same form as in
\p{QSU3} where now
\be
\lb{ASp4}
{\cal A}^{(a)}  = - \frac i2 \left(  \frac {\bar z^{(a)}}
{\bar z^{(a)}  z^{(a)} + m^2} +  \frac {\bar z^{(b+a)}}
{\bar z^{(b+a)}  z^{(b+a)} + m^2} +  \frac {2 \bar z^{(b+2a)}}
{\bar z^{(b+2a)}  z^{(b+2a)} + m^2}\right), \nn 
 {\cal A}^{(b)}  = - \frac i2 \left(  \frac {\bar z^{(b)}}
{\bar z^{(b)}  z^{(b)} + m^2} +  \frac {\bar z^{(b+a)}}
{\bar z^{(b+a)}  z^{(b+a)} + m^2} +  \frac { \bar z^{(b+2a)}}
{\bar z^{(b+2a)}  z^{(b+2a)} + m^2}\right)
  \ee
Note the appearance of the coefficient 2 in the last term in ${\cal A}^{(a)}$. This corresponds to the 
coefficient 2 with which the simple coroot $a$ enters in the expansion of the coroot $b+2a$.
The magnetic field determinant is 
 \be
\lb{detBSp4}
 \left\| 
\begin{array}{cc} {\cal B}_a +  {\cal B}_{b+a} + 4{\cal B}_{b+2a} & {\cal B}_{b+a} + 
2{\cal B}_{b+2a} \\
{\cal B}_{b+a} + 2 {\cal B}_{b+2a} & {\cal B}_{b} + {\cal B}_{b+a} + {\cal B}_{b+2a}  
\end{array} \right\| = \nn
{\cal B}_a ( {\cal B}_b + {\cal B}_{b+a} +   {\cal B}_{b+2a} ) + 
 {\cal B}_{b+a}  ({\cal B}_{b} +  {\cal B}_{b+2a} ) +   4 {\cal B}_b {\cal B}_{b+2a} \, .
  \ee
The integral $\left(\int  {\cal B}_{a} {\cal B}_{b} \right)/(4\pi^2)$ is equal to 1. Four other integrals
of the products ${\cal B}_{a} {\cal B}_{b+2a}$,  ${\cal B}_{b} {\cal B}_{b+a}$, 
${\cal B}_{a} {\cal B}_{b+2a}$, and ${\cal B}_{a} {\cal B}_{b+a}$ are reduced to $\int 
{\cal B}_{a} {\cal B}_{b}$ by the variable change with a unit Jacobian and also give 1. 
On the other hand, the integral 
$\left(\int  {\cal B}_{b} {\cal B}_{b+2a} \right)/(4\pi^2)$ is equal to 1/4.
We thus obtain the result
 \be
\lb{ISp4}
 I = 6
 \ee
for the index integral.

{\it c)} $G_2$.
The positive coroots of $G_2$ are depicted in Fig.3. The (regularized)
 supercharge has, again, the form \p{QSU3} with 
\be
\lb{AG2}
{\cal A}_a  = - \frac i2 \left(  \frac {\bar z^{(a)}}
{\bar z^{(a)}  z^{(a)} + m^2} +  \frac {\bar z^{(b+a)}}
{\bar z^{(b+a)}  z^{(b+a)} + m^2} +  \frac {2 \bar z^{(b+2a)}}
{\bar z^{(b+2a)}  z^{(b+2a)} + m^2} \right. \nn
\left.  + \frac {3 \bar z^{(b+3a)}}
{\bar z^{(b+3a)}  z^{(b+3a)} + m^2}  + 
\frac {3 \bar z^{(2b+3a)}}
{\bar z^{(2b+3a)}  z^{(2b+3a)} + m^2} \right)\, , \nn 
 {\cal A}_b  = - \frac i2 \left(  \frac {\bar z^{(b)}}
{\bar z^{(b)}  z^{(b)} + m^2} +  \frac {\bar z^{(b+a)}}
{\bar z^{(b+a)}  z^{(b+a)} + m^2} +  \frac { \bar z^{(b+2a)}}
{\bar z^{(b+2a)}  z^{(b+2a)} + m^2}  \right. \nn
\left. +  \frac { \bar z^{(b+3a)}}
{\bar z^{(b+3a)}  z^{(b+3a)} + m^2} + \frac {2 \bar z^{(2b+3a)}}
{\bar z^{(2b+3a)}  z^{(2b+3a)} + m^2} \right)
  \ee 
The matrix of magnetic fields is
 \be
\lb{BG2}
{\cal B}_{aa} &=& {\cal B}_a +  {\cal B}_{b+a} + 4{\cal B}_{b+2a} + 
9 {\cal B}_{b+3a}
+ 9 {\cal B}_{2b+3a} \, , \nn
{\cal B}_{ab} &=& {\cal B}_{ba} = {\cal B}_{b+a} + 2 {\cal B}_{b+2a} + 
3 {\cal B}_{b+3a} + 6 {\cal B}_{2b+3a} \, , \nn
 {\cal B}_{bb} &=& {\cal B}_{b} + {\cal B}_{b+a} + 
{\cal B}_{b+2a} + {\cal B}_{b+3a}+
4{\cal B}_{2b+3a} \, .  
    \ee
As in the all cases considered above, only the products of magnetic fields 
corresponding to different coroots survive in the determinant. 
\begin{enumerate}
\item There are 27 products of the type ${\cal B}_{b} {\cal B}_{b+3a}$ 
where both coroots are long, each such product entering
 with the Jacobian factor $1/9$.
 \item There are 3 products of the type ${\cal B}_{a} {\cal B}_{b+a}$ 
 when both coroots are short. They enter with a unit factor.
 \item There are altogether 18 products when one of the 
coroots is long and another short. 12 such products  of the type 
 ${\cal B}_{a} {\cal B}_{2b+3a}$ correspond to orthogonal coroots, they enter 
with the   Jacobian factor $1/4$. 6 other long-short products are of the type
 ${\cal B}_{a} {\cal B}_{2b+3a}$. They enter with the unit Jacobian.
 \end{enumerate}
 Adding all together, we obtain
 \be
\lb{IG2}
I = \frac {27}9 + 3 + \frac{12}4 + 6 = 15 \, . 
 \ee

\subsection{Strings on the maximal tori.}
To make the spectrum discrete and the index well-defined, we have to compactify 
${\mathcal C}^r \to T_G \times T_G$ where $T_G$ is the maximal torus of the corresponding group. 
This amounts to solving the problem for a regular lattice of strings.

\vspace{1mm}

{\it (a) } $SU(N)$.

Consider $SU(3)$ first. Consider the lattice of $SU(3)$ Dirac strings placed 
at the nodes of the lattice in Fig. \ref{SU3lattice}. 
As a first try, 
we replace \p{ASU3prost} by the  sums

\be
\lb{ASU3lattice}
 {\cal A}^{(a)}  =  - \frac i2 \sum_{pq} \left(   \frac {\bar z_{pq}^{(a)}}
{\bar z^{(a)}_{pq}  z^{(a)}_{pq} + m^2} +   \frac {\bar z^{(a+b)}_{pq}} 
{\bar z^{(a+b)}_{pq}  z^{(a+b)}_{pq} + m^2}  \right) \, ,  \nn
 {\cal A}^{(b)} =   - \frac i2 \sum_{pq} \left(   \frac {\bar z^{(b)}_{pq}}
{\bar z^{(b)}_{pq}  z^{(b)}_{pq} + m^2} +   \frac {\bar z^{(a+b)}_{pq}} 
{\bar z^{(a+b)}_{pq}  z^{(a+b)}_{pq} + m^2}  \right) \, ,
 \ee
where $z_{pq}^{(a)} = (\vecb{z} + p\vecb{a} + q \vecb{b}) \cdot \vecb{a}$, 
etc with complex integer  $p,q$. 

Several remarks are in order here, however. 

First of all, one should not sum over {\it all} $p,q$. 
Such sum would be infinite by two different reasons:
    \begin{enumerate}
\item The projections  $z^{(a)}_{pq}$ etc depend not on  $p,q$ separately, but on their certain
combinations: $2p-q$ for $z^{(a)}_{pq}$,  $2q-p$ for $z^{(b)}_{pq}$, and  $p+q$ for $z^{(a+b)}_{pq}$. To avoid 
a $\infty^2$-fold  counting, one should only sum over {\it these}  parameters:
 \be
\lb{factorsum} 
\sum_{pq}   \frac {\bar z_{pq}^{(a)}}
{\bar z^{(a)}_{pq}  z^{(a)}_{pq} + m^2}  \ \to \ \sum_{2p-q}'  \frac {\bar z_{pq}^{(a)}}
{\bar z^{(a)}_{pq}  z^{(a)}_{pq} + m^2} \, , 
  \ee
etc.
  
 \item A {\it naive} sum in the R.H.S. of \p{factorsum} still diverges. The prime put there means 
subtraction of a certain linear function $\vecg{\alpha} \vecb{z} + \beta$ with infinite coefficients, like
in \p{zeta}.
 \end{enumerate}

Second, one can be easily convinced that the potentials \p{ASU3lattice} are  
singular not only at the nodes of the lattice, but also at the fundamental coweights points \p{kovesa} translated with 
$p\vecb{a} + q\vecb{b}$. The torus $T_{SU(3)} \times T_{SU(3)}$ involves 9 such points. In other words, one could
consider right from the beginning the sums  \p{ASU3lattice} with  
$z_{pq}^{(a)} = (\vecb{z} + p\triangle\!\!\!\!\!\triangle + q\Box\!\!\!\!\!\Box) \cdot \vecb{a}$, etc.

These potentials enter the supercharge \p{QSU3} and the corresponding Hamiltonian. 
The wave functions satisfy the quasiperiodic boundary conditions with the same topology 
as the boundary
conditions \p{bc} with $k=3$. 
As was the case for $SU(2)$, an asymmetric choice $\alpha_a(\vecb{x}, \vecb{y}) = 0,\
 \beta_a(\vecb{x}, \vecb{y}) = 12\pi \vecb{a} \vecb{x}$ is more convenient
for us. 
 The spectrum is now discrete. There are $3 \cdot 9 = 27$   vacuum wave functions satisfying 
$Q \Psi = 0$ and these boundary conditions. In the massless limit, they acquire the form
  \be
\lb{vacSU3torus}
| 0 \rangle_{SU(3) \ {\rm lattice}}  \ =\  
\sqrt{ \frac {\Psi ( \vecb{ \bar z})}  
{ \Pi^{SU(3)} (\vecb{z})}}
  \ee
with  $\Psi (\vecb{z})$ being one of 27 theta functions satisfying the boundary conditions \p{bcSU3} with $k=3$. 
 
The index of such system is thus equal to 27. Note that all eigenfunctions \p{vacSU3torus} are singular
 (for nonzero mass, the singularity is smeared out).

It is the  potentials \p{ASU3lattice} that are generated by the gluon loops in SYMCS with $SU(3)$ gauge group. 
However, one can also construct a less dense string lattice where the $SU(3)$ 
Dirac strings are placed {\it only} at the black blobes in Fig.1 and not at the points \p{kovesa}.
 To this end, one should start not with the superpotential \p{Wanyreg}, but replace $z^p$ there by the fundamental
weights (with the factor $\frac {\sqrt{3}}2$), $z^8, \ \frac {z^8}2 \pm \frac {\sqrt{3} z^3}2$. 
The corresponding potentials have the same form 
as in \p{calAa}, but with $z^3$ and $z^8$ interchanged. 
\footnote{An alternative would be to stay with \p{calAa}, but translate it over the lattice rotated by 
$\pi/2$ compared to Fig. 1.}

In this case, there is only one string in $T_{SU(3)} \times T_{SU(3)}$ and there are only three vacuum states. 
In the massless limit, their wave functions have the form 
  \be
\lb{SU3redkijtor}
| 0 \rangle_{{\rm spacy} \ SU(3) \ {\rm lattice}}  \ =\  
\sqrt{ \frac {\Psi_{0, \triangle\!\!\!\!\!\triangle, \Box\!\!\!\!\!\Box}  ( \vecb{ \bar z})}  
{ \Theta^{SU(3)} (\vecb{z})}}
  \ee
 (see \p{psi3k1}, \p{TetSU3}).

A generalization to $SU(N)$ is straightforward. 
One can  either place the properly defined $SU(N)$ strings at the nodes of the coroot lattice, 
in which case the index is $N^{N-2}$, or place them also at the fundamental coweights points giving $I = N^N$. 

\vspace{1mm}
{\it b)} $Sp(4)$.

As was mentioned after \p{PiSp4}, the maximal torus of $Sp(4)$ involves a special point, 
$\vecb{w}_1 = \vecb{a} + \vecb{b}/2$. This is a fundamental coweight
corresponding to the element  $-1\!\!\!\!1$ of the center $Z_2$ of $Sp(4)$. Bearing in mind further applications
to SYMCS, consider the lattice of strings generated by the vectors $\vecb{w}_1$ and $\vecb{w}_2 = \vecb{b}/2$. 
Each node of this lattice [there are four such nodes in $T_{Sp(4)} \times  T_{Sp(4)} $]
 contributes 6 to the index.
  However, the fields
\be
\lb{ASp4lattice}
{\cal A}^{(a)}  = - \frac i2 \sum_{pq}' \left(  \frac {\bar z^{(a)}_{pq}}
{\bar z^{(a)}_{pq}  z^{(a)}_{pq} + m^2} +  \frac {\bar z^{(b+a)}_{pq}}
{\bar z^{(b+a)}_{pq}  z^{(b+a)}_{pq} + m^2} +  \frac {2 \bar z^{(b+2a)}_{pq}}
{\bar z^{(b+2a)}_{pq}  z^{(b+2a)}_{pq} + m^2}\right), \nn 
 {\cal A}^{(b)}  = - \frac i2 \sum_{pq}' \left(  \frac {\bar z^{(b)}_{pq}}
{\bar z^{(b)}_{pq}  z^{(b)}_{pq} + m^2} +  \frac {\bar z^{(b+a)}_{pq}}
{\bar z^{(b+a)}_{pq}  z^{(b+a)}_{pq} + m^2} +  \frac { \bar z^{(b+2a)}_{pq}}
{\bar z^{(b+2a)}_{pq}  z^{(b+2a)}_{pq} + m^2}\right)
  \ee
where $z_{pq}^{(a)} = (\vecb{z} + p\vecb{w}_1 + q \vecb{w_2}) \cdot \vecb{a}$, etc, and the symbol $\sum_{pq}'$ 
has the same meaning as in \p{factorsum} [that is, for example, the first sum in the first line in 
\p{ASp4lattice} is actually the sum over the complex integer $p-q$ regularized as in \p{zeta}] 
are both singular 
in the massless limit not only at the nodes of this lattice, but also at the points \p{12zeroes} 
(with $p\vecb{w}_1
+ q\vecb{w}_2$ added). 

Consider one of these points, $\vecb{z_*} = \left(\frac 12, 0\right)$. 
In this case, the projections $\vecb{z}^{(b)}_{pq}$ and $\vecb{z}^{(b+2a)}_{pq}$ 
associated with the long coroots are all nonzero
and the corresponding contributions to the vector potentials \p{ASp4lattice} are not singular. 
One can disregard them. 
On the other hand, the short coroot projections 
$\vecb{z}^{(b+a)}_{* \ pq}$ vanish when $p+q = 0$ and
   $\vecb{z}_{* \ pq}^{(a)}$ vanish when $p-q = -1$.

It is convenient to pose now  $\vecb{z} = \vecb{z_*} + \vecb{\delta}$ and 
express the supercharge in terms of $\delta^{(a)} = \delta_1$ and $\delta^{(a+b)} = \delta_2$, 
and similarly for $\psi$. 
  Then at the vicinity of $\vecb{z_*}$, one can neglect the nonsingular contributions due to the long coroots 
$\vecb{b}$, $\vecb{b} + 2\vecb{a}$, and the supercharge acquires the form
  \be
\lb{Qnovpole}
Q = -i\sqrt{2} \left[ \psi_1 \left( \frac \partial {\partial \delta_1} + \frac 1 {2\delta_1} \right)
+   \psi_2 \left( \frac \partial {\partial \delta_2} + \frac 1 {2\delta_2} \right) \right] \, .
  \ee
In other words, the supercharge represents a sum of two $SU(2)$ supercharges and the Hamiltonian ---
the sum of two $SU(2)$ Hamiltonians. There is only one zero mode representing the product of
the zero modes \p{zeromodetorus}. The same is true for 11 other special points in \p{12zeroes}.

The net index is equal to 
 \be
\lb{I36}
I_{Sp(4)}^{\rm torus} \ =\ 4\cdot 6 + 12 = 36 \, .
 \ee 

The result \p{I36} refers to the lattice of strings generated by the 
vectors $\vecb{w}_1$,  $\vecb{w}_2$. For the coroot 
lattice generated by $\vecb{a}$ and $\vecb{b}$,  $T_{Sp(4)} \times  T_{Sp(4)} $ involves 
just one genuine $Sp(4)$ string while the points $\vecb{z} = \vecb{w}_1, i\vecb{w}_1, (1+i) \vecb{w}_1$ involve, in contrast to what was the case for \p{ASU3lattice},
 only the 'tensor products'' of two independent $SU(2)$ strings and contribute 1 to the index.
 We obtain
 \be
\lb{I9}
I_{Sp(4)}^{\rm coroot\ lattice} \ =\  6 + 3 = 9 \, .
 \ee

\vspace{1mm}
{\it c)} $G_2$.

$G_2$ has no center and the strings \p{AG2} are located only at the nodes of the coroot lattice 
in Fig. \ref{G2lattice},
$\vecb{z} \to \vecb{z} + p \vecb{a} + q \vecb{b}$. There is only one such node in $T_{G_2} \times
T_{G_2}$ and this contributes 15 to the index. 

 Besides, there are 8 special points \p{8zeroesG2}
where the function \p{PiG2} has  the zeroes of the 3-d order and simultaneously the vector potentials
\p{AG2} translated over the lattice as in \p{ASU3lattice} and \p{ASp4lattice} are both singular.
Consider for example the pole at $\vecb{z}_* = \triangle\!\!\!\!\!\triangle \ =\ \vecb{a} + \vecb{b}/3$
($\vecb{b}$ being the coroot of $G_2$).  It is not difficult to see that
the only projections that vanish there are $z^{(a)}_{pq}$ with $2p - 3q  = -1$, $z^{(b+a)}_{pq}$ with $-p+3q = 0$, and
$z^{(b+2a)}_{pq}$  with $p = -1$. They are all associated with
the short coroots.

It follows then that, at the vicinity of $\vecb{z}_*$ [and also at the vicinity of 7 other poles \p{8zeroesG2}], 
the gauge fields 
and the supercharge have the same form as for $SU(3)$. In other words, the contribution of 
each such pole to the index is 3.  

Finally, the lattice sums $\sum_{pq} {\cal A}^{(a,b)}_{pq} $ are singular at nine points 
\p{9zeroesG2}.
The associated singularities are ``simple'',  like in  \p{Qnovpole}. Each such singularity contributes 1 
to the index. 
We finally obtain
 \be
\lb{I48}
 I_{G_2}^{\rm torus} \ =\ 15 + 3 \cdot 8 + 9 \ =\ 48 \, . 
 \ee

\subsection{Index of SYMCS.}

{\it a)} $SU(2)$. 

\vspace{1mm}

The tree-level effective Hamiltonian describes the motion in a 
homogeneous magnetic field of flux $2k$. As was mentioned in the Introduction, gluon 
loops bring about the flux lines placed at four corners 
 \be
\lb{corners}
z = 0, \frac 12, \frac i2,  \frac {1+i}2 \, ,
  \ee
each line carrying the flux $+1$. The extra net flux $+4$ brings about 4 extra states 
in the Hamiltonian. However, as was mentioned above, these extra states become singular 
in the massless limit and should be disregarded. If we keep mass finite, 
these states are not singular [cf. \p{zeromode}], but they have an essential support in the
region $|z - z_{\rm pole}| \sim m$ where the Born-Oppenheimer approximation (on the basis of which 
the whole method is based) does not apply, and these extra states sitting in the string cores 
should be disregarded by that reason \cite{ja2,ja3}. 

There are also fermion loops bringing extra flux lines at the  corners \p{corners}, 
but the flux of each fermion-induced line is $-1/2$. The Schr\"odinger
problem in the field of an {\it individual} fractional flux line is ill-defined, 
but when there are four such lines, the net flux is integer and the explicit expressions for 
the vacuum BO functions can be written. They have the form
  \be
\lb{chieffSU2}
 \chi^{\rm eff}_m(z, \bar z) \sim e^{-\pi k \bar z z} e^{\pi k \bar z^2} \, 
Q^{2k-2}_m(\bar z) \Pi^{3/4}(\bar z) \Pi^{-1/4} (z) 
 \ee
with the functions $Q^q_m$, $\Pi$ defined in \p{Qqm}, \p{Pidef}. The fractional powers of $\Pi$ are
due to fractional fluxes, but the function \p{chieffSU2} is uniquely defined and vanishes at the
corners. The parameter $m$ (nothing to do with the mass!) changes from 1 to $2k-2$ and we see 
thus $2k-2$ vacuum states.
\footnote{This counting coincides with the heuristic 
\be
{\rm relevant \ flux}  \ =\ (2k)_{\rm tree} - 4 \cdot \left( \frac 12 \right)_{\rm ferm. \ loops}
 = 2k-2  \, ,
 \ee
with the contribution of the gluon loops disregarded.}      

Let us remind how \p{chieffSU2} is derived \cite{ja3}. Each corner \p{corners} carries the flux $1-1/2 = 1/2$;
In each such corner, for example near the origin  in the region $m \ll |z| \ll 1$, the effective wave function satisfies
the equation 
  \be
  \lb{eqchi} 
   \left( \frac \partial {\partial z} + \frac 1 {4z} \right) \chi^{\rm eff} \ =\ 0
  \ee
and hence involves the factor $\sim z^{-1/4}$. The effective wave function that behaves in such a way in the vicinity
of each corner \p{corners} and satisfies the proper twisted boundary conditions corresponding to the net flux $2k+2$
has the form
 \be
 \lb{2k+2}
 \chi \ \propto \ \frac {Q^{2k+2}_m(\bar z) }{[\Pi(z) \Pi(\bar z)]^{1/4}} \, .
 \ee
The requirement for the wave function to be regular at the corners implies that $Q^{2k+2}_m(\bar z)$ has zeroes there
and this means that it can be represented as $\Pi(\bar z) Q^{2k-2}_m(\bar z)$. Which brings us to \p{chieffSU2}, where
the exponential factors are the same as in the tree--level wave functions. The first factor takes its origin in the
constant part of the magnetic field. The second factor (together with the first one and with other factors) makes
the probability density $\sim |\chi^{\rm eff}|^2$ periodic.   

Note that the actual number of vacuum SYMCS states is, however, less than $2k-2$ because we have 
to impose an additional constraint for the states to be gauge-invariant,
 which entails Weyl-invariance of the effective wave functions. For $SU(2)$, this means invariance
under $z \to -z, \bar z \to -\bar z$. This leaves only $(k-1)+1 = k$ states, in accordance with
\p{Ik2}.  

\vspace{1mm}

{\it b) $SU(N)$ , $Sp(4)$, $G_2$ .}

 \vspace{1mm}

Consider $SU(3)$ first. The gluon loop corrections generate  Dirac strings \p{calAa} at 9 points \p{kovesa}.
 The fermion loops generate there {\it fractional}  strings
with the vector potentials involving an additional factor $1/2$.  Similarly to what was mentioned for the 
$SU(2)$ case, the Schr\"odinger problem in the field of an individual fractional string is ill--defined. However, 
it is {\it well} defined when there are 9 such strings and, on top of that, also the constant magnetic field
corresponding to a half-integer level $k$. In the full analogy with \p{chieffSU2}, 
one can derive  
   
\be
\lb{chieffSU3}
 \chi^{\rm eff}_{SU(3)} (\vecb{z},  \vecb{\bar z}) \propto 
 \left[\Pi^{SU(3)}( \vecb{\bar z}) \right]^{3/4}
 \left[ \Pi^{SU(3)} (\vecb{z}) \right]^{-1/4} \, 
\theta^{SU(3)} ( \vecb{\bar z}) \, , 
 \ee
 where $ \theta^{SU(3)} ( \vecb{\bar z})$ is a theta function satisfying the boundary conditions \p{bcSU3}
 with $k_{\rm eff} = k - \frac 32$ [cf. \p{kren} !] . This gives 
 \be
 \lb{hrenSU3}
I = 3\left(k - \frac 32 \right)^2
 \ee 
states and, after imposing the 
Weyl invariance condition, results in  $\frac {k^2 - 1/4}2$ vacuum states in SYMCS theory, in accordance
with \p{Ik3}. 

Note that the counting \p{hrenSU3} has a relationship to the value of the 
index, $I=27$, for the lattice of $SU(3)$ Dirac strings including fundamental coweights, which was evaluated 
earlier. Indeed, disregard in \p{hrenSU3} the constant field (set $k=0$) and replace $3/2$ by 3 (go over from
the fermion-induced fractional strings to gluon-induced integer strings). We obtain $I = 27$.    

In a similar way, we obtain 
\be
 \lb{hrenSUN}
I = N\left(k - \frac N2 \right)^{N-1}
 \ee 
``pre-Weyl'' states for $SU(N)$, which leads to \p{IkN}. 

Let us discuss $Sp(4)$. The effective wave function has exactly the same form as in \p{chieffSU3}
where one should replace the functions \p{PiSU3} by the functions \p{PiSp4}, while the function
$\theta^{Sp(4)} ( \vecb{\bar z}) $ satisfies the boundary conditions \p{bcSp4} with  
$k_{\rm eff} = k - \frac 32$ (the Casimir eigenvalues for $SU(3)$ and $Sp(4)$ are the same).
 This gives 
  \be
 \lb{hrenSp4}
I = 4\left(k - \frac 32 \right)^2
 \ee
  pre-Weyl states. If setting $k=0$ and replacing $3/2 \to 3$, we obtain 36 states, which conforms
with the counting \p{I36}. 
When Weyl invariance requirement is imposed, only $\frac {k^2 - 1/4}2$ states is left 
\footnote{This number is obtained setting $r=2$ in the general 
tree-level result \cite{ja1} for $Sp(2r)$, 
 \be
\lb{ISpr}
 I = \ \left(  \begin{array}{c} k+r \\ r \end{array} \right) 
  \ee
and replacing $k \to k-3/2$.},  the same number as for $SU(3)$.

The  effective wave functions for $G_2$ have a similar form. The difference with $SU(3)$ is that the tree value
of $k$ is shifted down not by $3/2$, but by $\frac 12 c_V[G_2] = 2$. This gives  
  \be
 \lb{hrenG2}
I = 3\left(k - 2 \right)^2
 \ee
pre-Weyl states.   If setting $k=0$ and replacing $2 \to 4$, we obtain 48 states, which conforms
with the counting \p{I48}. After Weyl invariance requirement is imposed, the final result for the index is
\cite{ja3}
   \be
 \label{indG2}
 I^{\rm tree}_{G_2}( k) \ =\ \left\{ 
                 \begin{array}{c}  \frac {(|k|+2)^2}4\  \ \ \ \ \ \ \ \ {\rm for\ even} \  k \\
                                   \frac {(|k|+1)(|k|+3)}4 \ \ \ \ \ \ \ \ {\rm for\ odd} \  k 
 \end{array}
               \right\} \ .
\ee

   The estimates \p{hrenSU3}, \p{hrenSUN}, \p{hrenSp4}, \p{hrenG2} confirm the recipe \p{recipe}.

\vspace{1mm}

{\it c) Other groups.}

 \vspace{1mm}

In addition to the groups discussed above, the SYMCS index can be easily evaluated also for higher symplectic groups.
At the tree level, the pre-Weyl counting for $Sp(2r)$ is $(2k)^r$. 
When Weyl invariance requirement is imposed, the result \p{ISpr} 
is obtained. When loop effects are taken into account, the reasoning above for $SU(2)$ and $SU(3)$ can be repeated
without change. We are led to the relation
      
\be
\lb{chieffSp2r}
 \chi^{\rm eff}_{Sp(2r)} (\vecb{z},  \vecb{\bar z}) \propto 
 \left[\Pi^{Sp(2r)}( \vecb{\bar z}) \right]^{3/4}
 \left[ \Pi^{Sp(2r)} (\vecb{z}) \right]^{-1/4} \, 
\theta^{Sp(2r)} ( \vecb{\bar z}) \, , 
 \ee 
where $\theta^{Sp(2r)} ( \vecb{ \bar z})$ is a theta function depending on $r$ complex arguments and satisfying
an obvious generalization of the boundary conditions \p{bcSp4} with  $k$ replaced by  $k - \frac{r+1}2$.
 This finally gives for positive $k$
\be
\lb{ISp2r}
I^{Sp(2r)} \ =\ \left( \begin{array}{c} k + \frac {r-1}2 \\ r \end{array} \right)\, , 
 \ee
and $I^{Sp(2r)} (-k) = (-1)^r\, I^{Sp(2r)}(k)$. 
  
For more complicated groups, the tree-level calculation has not been performed yet, but the $SU(2)$ reasoning
regarding a proper account of loop corrections {\it can} be generalized for an arbitrary group. The recipe \p{recipe}
is thus confirmed even though the L.H.S. and the R.H.S. of this relation are not yet known in a general case. 

For $SU(N)$, $Sp(4)$, and $G_2$, we also observed that the pre-Weyl SYMCS index counting matches well the evaluations
of the index of the string lattices performed in Sect. 4,3. One can {\it conjecture} that this matching works also for 
more complicated groups, but an {\it explicit} proof of this statement is not so easy. We have seen that, 
for non-unitary groups, the counting expected on the basis of the SYMCS analysis  is reproduced after 
adding {\it different} 
nontrivial contributions to the index [see Eqs.\p{I36} and \p{I48}]. 

It would be nice to explore this interesting conspiracy for higher orthogonal and exceptional groups.

\section{Discussion.}

Our motivation to perform this study was the wish to prove more or less rigourously the assertion \p{recipe} 
for the groups of higher rank. We believe that this goal has now been achieved.

However, a byproduct of such a study --- an analysis of generalized multidimensional Dirac strings might prove to be
also interesting, even more interesting than this anticipated result. The index integrals associated with these generalized
strings are nontrivial. Even for $SU(N)$, we were able to explicitly calculate it only for $N \leq 6$. 
The result \p{InumbersN} is a conjecture. In other cases, we performed this calculation only for the groups of rank 2.

It would be interesting to calculate these integrals also for other groups. Physics applications disregarding, each such integral
represents an integer number associated with a given group. What is a mathematical nature of this number ? How can it be explained ?
This question is akin to the question of the so called {\it principal contribution}   to the index in maximally supersymmetric
gauge matrix models. For a given simple group, 
the principal contribution to the index 
represents a bizarre fractional number (see \cite{Kac2}, \cite{Stau} and references therein) whose mathematical nature is now obscure.

The problem of calculating the index for a lattice of such generalized Dirac strings proved to be also rather nontrivial.
Here the final result was anticipated on the base of the SYMCS analysis. However, for non-unitary groups, 
it was obtained 
as a sum of different nontrivial contributions. This reminds the story for the Witten index of pure SYM theory in 
four dimensions. The physical arguments
provide a universal answer: $I = c_V \equiv h^\vee$ for any gauge group. However, to reproduce this result for 
higher orthogonal and exceptional groups in the framework of BO analysis 
turned out to be a highly nontrivial task, and it took almost 20 years to  finally resolve it \cite{Kac1}, 
\cite{Keur}. The value $c_V$ is obtained in these cases as a sum of different nontrivial contributions.

One can also recall the problem of index evaluation for  SYMCS theories with matter. 
In  \cite{ja4}, we verified in 
 some special cases the generic conjecture of \cite{IntSei} for the index of a SYMCS theory with $SU(N)$ gauge group
involving also  matter multiplets in different representations. Also in this case one can 
observe a kind of conspiracy: a simple result is obtained as a sum of complicated individual terms. 

It would be rather desirable to achieve a better understanding of all these conspiracies.

\section*{Appendix. Proof 
\footnote{We constructed it together with 
J.-L. Milhorat whose aid I appreciate.}
of the relation \p{sumkvadratov}.}

\setcounter{equation}0
\def\theequation{A.\arabic{equation}}

The Casimir operator is defined as 
\be
\lb{defCasimir}
\hat{\cal C} x \ =\ [T^a, [T^a, x] ] \, ,
  \ee
where $T^a$ represent a particular orthogonal basis in the Lie algebra normalized with respect to the Killing form
$\langle x, y \rangle$ normalized such that $\langle \alpha^\vee, \alpha^\vee \rangle  = 4$ for {\it short} coroots. 
This corresponds to the normalization $\langle \alpha, \alpha \rangle $ for the long roots. 
For example, for $su(N)$ with
$x$ being the Hermitian $N \times N$ matrices, this Killing form coincides with 2Tr$\{xy\}$. 
Note the difference by the 
factor 4 with the convention \p{nashKilling} used throughout the text ! 
It is known that the operator \p{defCasimir} is proportional
to the unit matrix.

Choose now the Chevalleu basis 
 \be
\lb{Chevalleu}
T^a = \left\{h^a, \frac {E_p + E_{-p}} {2\sqrt{ d_p} },\  \frac {i(E_{-p} - E_p)} {2 \sqrt{d_p} }  \right\} \, ,
 \ee 
where $h^a$ 
belong to the Cartan subalgebra, $E_{\pm p}$ 
are positive and negative root vectors normalized such that $[E_p, E_{-p}] = \alpha_p^\vee$,  $d_p = 1$ 
for short coroots and $d_p = 2$ or $d_p = 3$ for long coroots.  Choose $x = h$ in the Cartan subalgebra.
 The Casimir operator \p{defCasimir} is then rewritten as
  \be
\lb{defCasimirviaroots}
\hat{\cal C} h  \ =\ \sum_p \frac 1{d_p} \, [E_p, [E_{-p}, h] ] \ =\ 
  \sum_p \frac 1{d_p} \, \alpha^\vee_p \alpha_p (h) \, .
  \ee
Projecting on $h$ with the generalization of the convention \p{nashKilling} (such that the length 
of the short coroots is normalized to unity) and using
    \be
  \lb{albetvee} 
  \langle \alpha_p^\vee, h \rangle \ =\ \frac {d_p}2 \alpha_p(h) \, ,
    \ee
we arrive at \p{sumkvadratov}.


\begin{thebibliography}{96}

\bibitem{Dunne} G.V. Dunne,  {\tt hep-th/9902115}. 

\bibitem{Wit99} E. Witten, in: 
   [Shifman, M.A. (ed.) {\it The many faces of the superworld} p.156] [{\tt hep-th/9903005}].

\bibitem{ja1}  A.V. Smilga,
{\it JHEP}  {\bf 1001} 086  (2010) [{\tt arXiv:0910.0803, hep-th}].

\bibitem{ja2}  A.V. Smilga,
{\it JHEP} {\bf 1205}  103 (2012) [{\tt arXiv:1202.6566, hep-th}].

\bibitem{ja3} A.V. Smilga, Uspekhi Fiz. Nauk, {\bf 184} (2014) 163 
[{\tt arXiv:1312.1804, hep-th}].


\bibitem{Wit82} E. Witten,  {\it Nucl. Phys. B} {\bf 202} 253  (1982).

\bibitem{DJT} S. Deser, R. Jackiw, and S. Templeton,  
{\it Ann. Phys. (NY)} {\bf 140}  372 (1982).

\bibitem{Novikov1} B.A. Dubrovin, I.M. Kri\v{c}ever, and S.P. Novikov,
 {\it Soviet Math. Dokl.} {\bf 229}  15 (1976).

\bibitem{Novikov2} 
B.A. Dubrovin and S.P. Novikov,  
{\it Soviet Phys. JETP} {\bf 79}  1006 (1980). 

\bibitem{flux} A.V. Smilga, J. Math. Phys. {\bf 53} 
(2012) 042103 [{\tt arXiv:1104.3986,  math-ph}].

\bibitem{Mumford} D. Mumford, 
{\it Tata Lectures on Theta} (Birkh\"auser Boston, 1983). 

\bibitem{Ryzhik} I.M. Ryzhik and I.S. Gradshteyn, 
{\it Table of integrals, series, and products}, 
(Elsevier/Academic Press, Amsterdam, 2007). 

\bibitem{Pi} K. Gawedzki and A. Kupiainen,
{\it Nucl. Phys. B} {\bf 320}  625 (1989).

\bibitem{Eli} S. Elitzur, G. Moore, A. Schwimmer, and N. Seiberg,
 {\it Nucl. Phys. B} {\bf  326}  108 (1989).

\bibitem{Laba}
J.M.F. Labastida and A.V. Ramallo,  {\it Phys. Lett. B} {\bf 227}  92 (1989). 


\bibitem{Kacpriv} V. G. Kac, private communication.

\bibitem{Hull} C.M. Hull,  {\tt hep-th/9910028}.

\bibitem{DiracSQM} E.A. Ivanov and A.V. Smilga, Int. J. of 
Mod. Phys. {\bf A27} (2012) 1250146
 [{\tt arXiv:1012.2069, hep-th}].

\bibitem{Kiskis} J. Kiskis, Phys. Rev. {\bf D15} (1977) 2329.

\bibitem{Wu} T.T. Wu and C.N. Yang, Nucl. Phys. {\bf B107} (1976) 365; 
S. Kim and C. Lee,
Ann. Phys. {\bf 296} (2002) 390  [{\tt  hep-th/0112120}].

\bibitem{Nambu}  
Y. Nambu, Nucl. Phys. {\bf B578} (2000) 590 [{\tt hep-th/9810182}].

\bibitem{Blok} 
        B.Yu. Blok and A.V. Smilga,  {\it Nucl. Phys. B} {\bf 287}  589 (1987).

\bibitem{Cecotti} S. Cecotti and L. Girardello, Phys. Lett. 
{\bf B110} (1982) 39.

\bibitem{Kac2} V.G. Kac and A.V. Smilga, Nucl. Phys.  {\bf B571} (2000) 515 

 [{\tt  hep-th/9908096}].

\bibitem{Stau} W. Krauth and M. Staudacher,  Nucl. Phys.  {\bf B584} (2000) 641

 [{\tt  hep-th/0004076}].

\bibitem{Kac1} V.G. Kac and A.V. Smilga,  in:
     [Shifman, M.A. (ed.) {\it The many faces of the superworld} p.185] 
[{\tt hep-th/9902029}].

\bibitem{Keur} A. Keurentjes,
{\it JHEP} {\bf 9905} (1999) 001; 014  [{\tt hep-th/9901154;   hep-th/9902186}] 

\bibitem{ja4} A.V. Smilga, Nuclear Physics B, in press [{\tt arXiv: 1308.5951, hep-th}].

\bibitem{IntSei}   K. Intriligator and N. Seiberg, JHEP {\bf 1307} (2013)  079 [{\tt arXiv: 1305.1633, hep-th}].







\end{thebibliography}
\end{document}